%% file: eth_review.tex
\documentclass[prx,twocolumn,showkeys,showpacs,superscriptaddress,nofootinbib,10pt
]{revtex4-1}
\usepackage{graphicx}  % needed for figures
\usepackage{grffile}
\usepackage{amsthm}
\usepackage{enumerate}
\usepackage{epsfig}
\usepackage{amsmath,amssymb,amsfonts,mathtools,bbm,MnSymbol,mathrsfs,xfrac}   % for math
\usepackage{longtable} %for tables
\usepackage{tabularx}
\usepackage[breaklinks=true,colorlinks=true,linkcolor=blue,urlcolor=blue,citecolor=blue]{hyperref}
\usepackage{comment}
\usepackage{color}
\usepackage[makeroom]{cancel}

\usepackage{soul}%for striking through text horizontally with \st{}
\renewcommand{\[}{\begin{equation}}
\renewcommand{\]}{\end{equation}}

\newcommand{\ket}[1]{|#1\rangle}

\newcommand{\braket}[2]{\langle#1|#2\rangle}
\newcommand{\pro}[2]{|#1\rangle\langle#2|}
\newcommand{\mean}[1]{\langle#1\rangle}

\newcommand{\ov}[1]{\overline{#1}}
\newcommand{\tr}{\mathrm{tr}}

\newcommand{\Eq}[1]{Eq.~\eqref{#1}}
\newcommand{\Sec}[1]{Sec.~\ref{#1}}
\newcommand{\Fig}[1]{Fig.~\ref{#1}}

\newcommand{\R}{{\hat{\rho}}}
\newcommand{\Ham}{{\hat{H}}}

\newcommand{\Ohat}{{\hat{O}}}

\definecolor{mygray}{gray}{0.6}
\theoremstyle{definition}

\newtheorem{conjecture}{Conjecture}

\begin{document}

\title{Eigenstate Thermalization Hypothesis}

\author{Joshua M. Deutsch}
\email{josh@ucsc.edu}
\affiliation{Department of Physics, University of California, Santa Cruz, CA 95064, USA}

\date{\today}

\begin{abstract} \input{abstract} \end{abstract}

\maketitle

\input{introduction}

\input{theproblem}

\input{routes}

\input{randommatrix}

\input{eth}

\input{numerics}

\input{coldatoms}

\input{conclusions}

\input{acknowledgments}

\bibliography{eth}
\end{document}

%% file: abstract.tex
The emergence of statistical mechanics for isolated classical systems comes about through chaotic
dynamics and ergodicity. Here we review how similar questions can be answered in quantum systems.
The crucial point is that individual energy eigenstates behave in many ways like a statistical
ensemble. A more detailed statement of this is named the Eigenstate Thermalization Hypothesis (ETH).
The reasons for why it works in so many cases are rooted in the early work of Wigner on random
matrix theory and our understanding of quantum chaos. The ETH has now been studied extensively by both
analytic and numerical means, and applied to a number of physical situations ranging from black hole
physics to condensed matter systems. It has recently become the focus of a number of experiments in
highly isolated systems. Current theoretical work also focuses on where the ETH breaks down leading to
new interesting phenomena. This review of the ETH takes a somewhat intuitive approach as to  why it
works and how this informs our understanding of many body quantum states.

%% file: introduction.tex
\section{Introduction}
\label{sec:introduction}

Over the last decade there has been a rapid growth in research studying the problem of
{\em thermalization} at a quantum level. Perhaps the first discussion of these issues started
with Schr\"odinger~\cite{schrodinger1927energieaustausch} and shortly after that, Von Neumann~\cite{von2010proof}
was able to make substantial headway into this deep and complex problem. 
Since then, there have been many approaches to understanding thermalization.
However the recent surge of interest has been focused on understanding thermalization from a microscopic
point of view, continuing the relentless campaign of physics to try to explain all phenomena 
from the Schroedinger equation. The path of getting from the microscopic to
the macroscopic is still not completely understood, but can be done when certain plausible assumptions are
introduced. The main one in this case is the ``Eigenstate Thermalization
Hypothesis" (ETH)~\cite{deutsch1991quantum,srednicki1994chaos,rigol2008thermalization}, 
and is the subject of this review. We will discuss the theoretical and numerical
evidence in support of the ETH, and also point out where it is known to fail.
This is not meant to be a comprehensive review of the field but a relatively
short and accessible introduction to readers interested in understanding
more qualitatively how the ETH comes about and how it is being currently studied.
For a more technical and comprehensive review, the reader is invited to peruse
Ref.~\cite{DAlessio2016quantum}.

\subsection{Why study the ETH?}

But why study the ETH and thermalization in general? The fact that a macroscopic body, such as a brick,
will ultimately come to equilibrium with its environment, seems so obvious from ones everyday
experience, that it might hardly seem of interest to pursue understanding why. But without such
experience, it is actually quite remarkable that this happens. Why should there be a way of defining an equilibrium
macroscopic state for a brick that does not depend, in detail, on its initial preparation? Physics tells us that
the evolution of a state depends completely on its initial conditions, and therefore the brick
should be described by $10^{23}$ numbers (actually $\exp(10^{23})$ in quantum mechanics), and not just a few.
And the fact that at a macroscopic level, a system's behavior becomes simple, 
means that things like memory
devices work reliably despite the fact that the quantum state of each device is completely different. 
The lack of dependence on the initial state, is what gives consistent behavior on a macroscopic scale, and
relaxation to an equilibrium thermal state.  Hence irreversibility and the 
second law of thermodynamics are closely related to thermalization. We will shortly review 
how thermalization in classical 
systems is closely related to idea of chaos and ergodicity. The ETH can be regarded, very broadly, as 
the quantum manifestation of such ergodic behavior.

Aside from understanding why things thermalize, the ETH sheds light on a system's behavior,
such as fluctuations and transport 
coefficients~\cite{srednicki1996thermal,srednicki1999approach,KhatamiPhysRevLett.111.050403,DAlessio2016quantum}.
It also helps to understand where systems fail to thermalize such as in 
``Many Body Localization"~\cite{AlthshulerPhysRevLett_78_2803,BaskO20061126,imbrie2016many,HusePhysRevB_88_014206,nandkishore2015many}, 
leading to the predictions and understanding of exotic new phenomena.

An example of a system very well described by a thermal state is a black hole.
Hence the quantum mechanical aspects of
chaos, thermalization and the ETH are very relevant to the understanding of the inner
workings of these elusive objects~\cite{marolf2013gauge}. 
Similarly thermalization is apparent in systems that are easier to study experimentally, such as
cold atoms\cite{levin2012ultracold,langen2015ultracold}, where many of 
the details of our picture of quantum thermalization can be tested. 

\subsection{Example of thermalization}

Now let us turn to a simple example of a system that illustrates the issues
involving thermalization that we discussed above.

\begin{figure}[]
\begin{center}
\includegraphics[width=1\hsize]{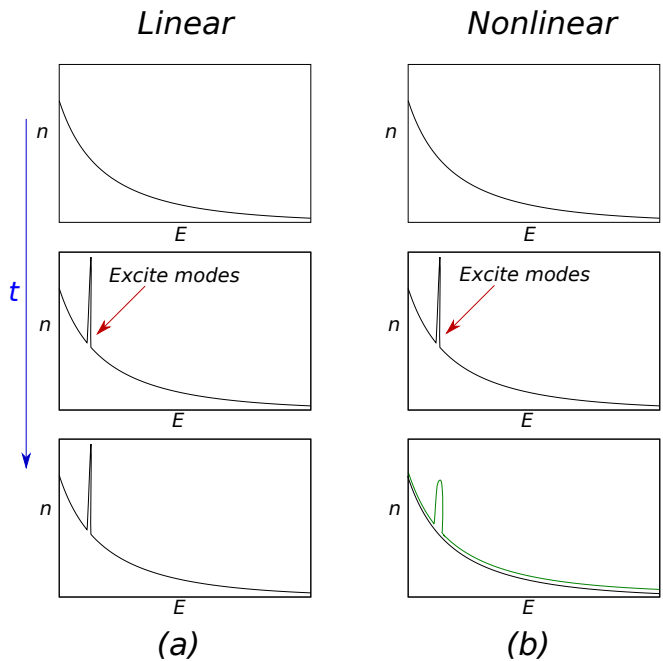}
\caption
{ 
(a) A linear system of phonons. Top panel: The average occupation of a mode at energy $E$ versus $E$ in thermal
equilibrium. Middle panel: Modes in a narrow band are excited. Lower panel: after long time, the
distribution remains unaltered. (b) The same situation with nonlinearity present. Lower panel:
The red curve, illustrates that the distribution at intermediate times as it relaxes back to the thermal 
result. The black curve shows the result after a long time, having relaxed back to the thermal
result. (The curves have been slightly displaced for clarity).
}
\label{fig:phonon_relaxation}
\end{center}
\end{figure}

Consider a perfectly harmonic crystal that is completely isolated. It is described classically by a linear
set of equations that can be diagonalized to yield different normal modes. Those modes
can then be quantized to give a complete quantum description of the system. 
First consider how we would reasonably expect the system to be  described in thermal equilibrium.
If we measured a single phonon property such as the occupation number $n(E)$ of a mode at any energy $E$, that should
be given by the Bose-Einstein distribution. To easily measure $n(E)$ experimentally, it would be better
to consider, instead, its average over all modes with similar energy, of which there are very many.
We then excite a range of modes closest to one energy $E_o$, say by optical means, and then remeasure $n(E)$
as a function of energy. This is illustrated in \Fig{fig:phonon_relaxation}. Because the occupation number for each mode 
in conserved, the probability distribution will have an extra peak at $E_o$
that will not change with time. The expected thermal distribution will never be attained. In this
sense, the system will not ``thermalize". The same problem occurs in a strictly classical treatment
of this problem

But such non-thermal behavior is not expected to occur in most experimental situations. We would
naturally expect, that in analogy with the classical case, that there will be an exchange of energy
with other modes, causing an eventual relaxation of the system to thermal equilibrium, with a
slightly higher temperature. How precisely this happens, even at the classical level, is not at all
obvious, and so it makes sense to first briefly review how classical systems manage to thermalize.
For a clear and more detailed exposition of the classical and a range of quantum approaches, the reader is invited
to read Ref. \cite{singh2013foundations}.

\subsection{Classical thermalization}

To understand thermalization, we confine ourselves for the moment, to a range of important quantities that
are at the heart of equilibrium statistical mechanics: equal time averages of observables. 
Consider some observable that varies as a function of time, $O(t)$, which could be, for example, the $z$ component of a
dipole moment, or the momentum of a single atom in a crystal. We would like to
find the time average of $O$, over an interval of time $\cal T$, that we will eventually extend to infinity.
In most situations, calculating the value of momentum over all time is essentially impossible,
and so it would seem that its time average would be as well. 

\begin{figure}[]
\begin{center}
\includegraphics[width=1\hsize]{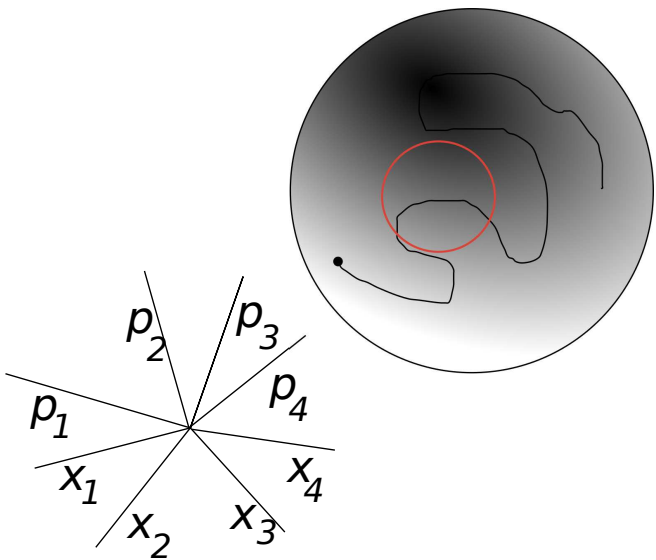}
\caption
{ 
A two dimensional projection of phase space with 8 canonically conjugate variables $q_1\dots q_4,p_1\dots p_4$.
The black line corresponds to the path of the system over some time interval. The sphere corresponds
to the points allowed by energy conservation. The red closed path corresponds to a periodic
trajectory.
}
\label{fig:phase_space}
\end{center}
\end{figure}

However there is a way of understanding this situation that makes answering such questions quite
manageable. We consider the phase space of a closed classical system, as illustrated in
\Fig{fig:phase_space}.

A complete description of the system is achieved by including all of the canonically conjugate coordinates. Say that there are
$N$ positional degrees of freedom $q_1\dots q_N$ and $N$ momenta $p_1\dots p_N$. We can view those
variables as a point in a $2N$ dimensional space, known as {\em phase space},
$\Gamma = \{q_1\dots q_N,p_1\dots p_N\}$ . As time progresses,
these coordinates will change as well, meaning that this point will move, as illustrated, tracing out a path.
Because energy is conserved, we know that this path will reside on a surface of constant energy.

In general, this path will be very complicated. If it succeeds in getting arbitrarily close to every point on this
surface, then the system is called ``ergodic"~\cite{farquhar1964ergodic,jancel2013foundations}. Ergodicity says, loosely speaking, that $\Gamma(t)$ will
get arbitrarily close to every point on the constant energy surface given a long enough time. 
We also know from Liouville's theorem, that the system will spend equal times in equal phase space 
volumes, so the trajectory will end up covering the ball uniformly. Such a path is shown in black in
\Fig{fig:phase_space}.

This means that instead of averaging an observable $O$ over time, $\langle O \rangle_t$, 
we could equally well average it over
phase space with the constraint that we are confined to this constant energy surface. That's a far
easier problem to calculate mathematically. More precisely, regarding the observable  as a function
of phase space $O(\Gamma)$, we can then say~\cite{ma1985statistical}
\[
\label{eq:O_T=int_O_over_int}
\langle O\rangle_t = \frac{\int_{S} O(\Gamma)d\Gamma}{\int_{S} d\Gamma}
\]
where we are integrating over a surface $S$ of constant energy.  And this is precisely how averages are described in the
classical ``microcanonical ensemble". It is also important to note that there other invariants, for
example total system momentum, that might also be conserved. In these cases, the surface must also
include these other invariants.

\begin{figure}[]
\begin{center}
\includegraphics[width=0.45\hsize]{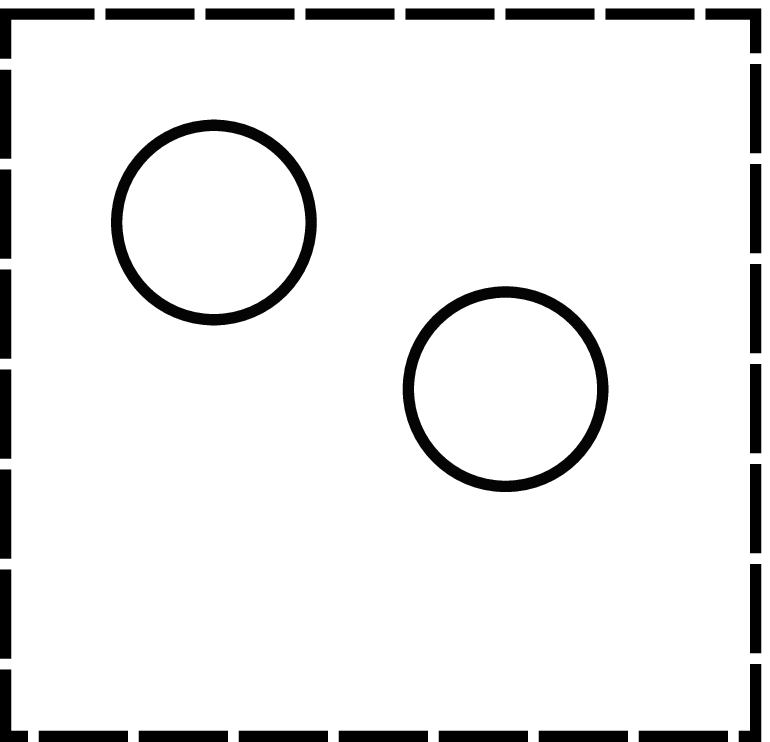}
\includegraphics[width=0.45\hsize]{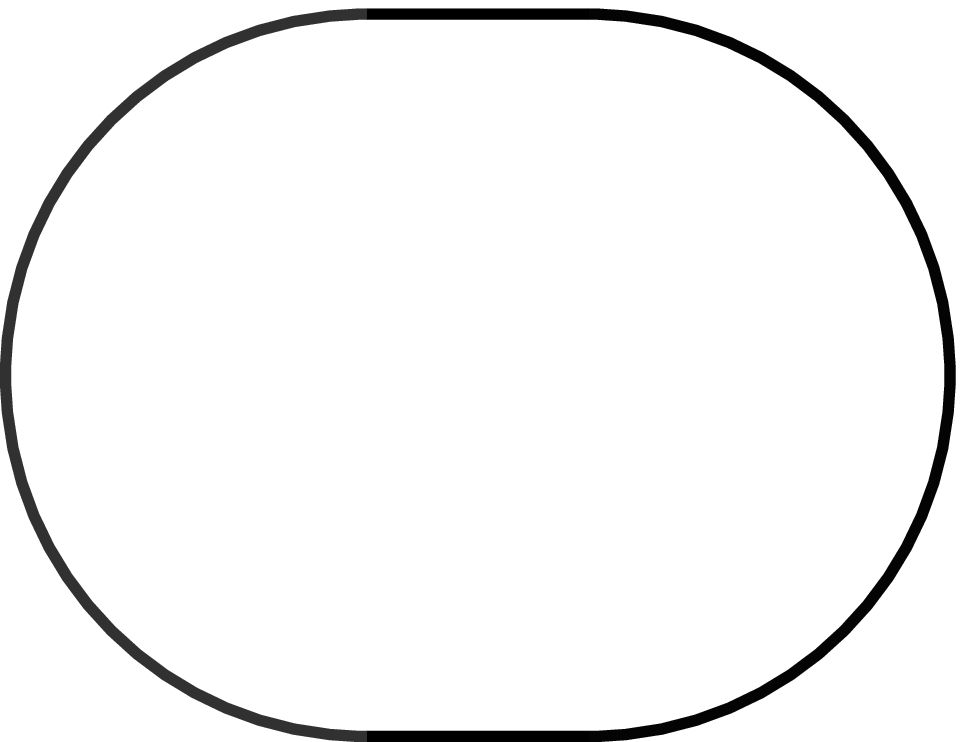}\\
(a)
\hskip 0.45\hsize
(b)
\caption
{ 
(a) Example of Sinai billiards: Two hard core spheres move in free space with periodic boundary conditions. 
(b) Bunimovich Stadium. A point particle moves in free space confined to a stadium with hard walls. The stadium
has semicircular sides and a straight mid-section.
}
\label{fig:SinaiBunimovich}
\end{center}
\end{figure}

In order for this to work, it would seem as if the system should be ergodic or at least close to it. 
There have been a few examples where it has been possible to prove ergodicity, most notably, a gas
of an arbitrary number of hard spheres in some volume~\cite{sinai1970dynamical,simanyi2004proof,simanyi2009conditional,szasz2008algebro}, 
often called ``Sinai Billiards" as illustrated in Fig. \ref{fig:SinaiBunimovich}(a). 
The proofs are quite involved\footnote{To be more precise, ergodicity has only been proven rigorously 
in some special cases that limit the number of spheres, 
or for systems where all of the masses are arbitrary, and then with the caveat that
the proof will not hold for a zero measure set of mass ratios\cite{simanyi2004proof,simanyi2009conditional,szasz2008algebro}.}, 
but the result tells
us that time averages are calculable through the microcanonical ensemble formula. 
Another such system that has been proved to be ergodic~\cite{bunimovich1979} is the ``Bunimovich Stadium", 
which describes the motion of a free particle inside a stadium
with hard walls that are circular on the sides, and straight in the middle, see
Fig. \ref{fig:SinaiBunimovich}(b).

But there are also other systems, the are ``integrable" where there are $N$ other invariants,
meaning that these are constants of motion. Such a closed trajectory is schematically represented by the red curve in 
\Fig{fig:phase_space}. Referencing our phonon example, the fact that the crystal does not 
thermalize is because of these extra invariants. In that case, each of these invariants is the energy of
a single normal mode. A typical trajectory of such a system is therefore described by the combined
motion of the normal modes, and will in general be quasi-periodic.
However, integrable systems are unusual, and not expected generically. For example,
any anharmonic term added, will make this problem non-integrable, or ``generic".

But in general, we do not expect that a generic classical system, for example a gas with Van-der Waals
interactions, or a model of phonons with anharmonic terms, will be, strictly
speaking, ergodic. For finite $N$, there has been a great deal of work on what happens in such systems. 
The Kolmogorov –- Arnold –- Moser (KAM) theorem tells us that for a weak anharmonic perturbation
of order $\epsilon$, most phase
space trajectories will continue to be quasi-periodic as in the integrable case. However as the strength
of the anharmonicity is increased, the fraction of such quasiperiodic trajectories is expected to
decrease. In real situations however we do not necessarily have very strong nonlinear terms in the
Hamiltonian, so why does statistical mechanics work in these cases?

What is generally believed is that as $N\rightarrow\infty$, the range of $\epsilon$'s where 
a significant fraction of quasiperiodic orbits survives becomes vanishingly small~\cite{falcioni1991ergodic}. 
Thus for an isolated system, for statistical mechanics to work, one needs to have large $N$.
In most experimental situations this is usually not an problem, because $N$ is normally very large
and therefore there will be a vanishingly small sets of initial
condition where the trajectories are quasiperiodic, and therefore the system can be considered
to be ergodic.

Related to ergodicity, is the idea of chaos. The idea is that two systems with slightly different initial
conditions will evolve into systems that are have very different coordinates, $\Gamma_1(t)$ and $\Gamma_2(t)$.
The rate of divergence can be characterized by
their Lyapunov exponents. For example, for two Sinai billiards in some closed volume,
the trajectories will be very sensitive to the initial conditions, and with every subsequent
collision, $\Gamma_1(t)$ and $\Gamma_2(t)$ will diverge from each other more strongly. However ergodicity is not equivalent
to chaos, a simple counter-example being a one dimensional harmonic oscillator, which is ergodic,
but close trajectories do not diverge from each other. Thus ergodicity ensures that phase space
is well stirred, but it does not ensure it is scrambled. However, for a large number of degrees of
freedom, we expect that most systems will be ergodic and chaotic.

But in some contexts, thermalization does imply scrambling, and more formally this idea
is called ``mixing"~\cite{arnol1968ergodic}. This can be thought of qualitatively to be related 
to mixing paints but in a high dimensional space. There are many definitions
for this, but it is says essentially understood as follows. Suppose we have an 
ensemble of systems with initial conditions in some
arbitrarily small region, analogous to a drop of dye in water. Then after a sufficiently long
time, the separate systems (analogous to dye molecules) will be spread uniformly over all of
accessible phase space. Thus such a system loses all correlations with its state at an earlier time.
Because of this, systems that are mixing will also be ergodic.

In practice, most large $N$ classical systems of physical relevance are strongly chaotic, so that they
are very close to being strongly mixing and ergodic. Later on we will contrast such systems with
well known quantum integrable systems.

\subsection{Quantum chaos}

Now we turn to trying to understand how chaos manifests itself in quantum mechanics. In fact, there is
some debate over the term ``Quantum Chaos" and Michael Berry prefers the term ``Quantum Chaology"~\cite{berry1987bakerian}
because as we shortly discuss, quantum mechanics cannot manifest chaos to the same extent that was described
above for classical mechanics. But there is definitely a set of phenomena that occur that
are related to classical chaos in many strongly interacting quantum systems.

\subsubsection{Random Matrices}

One of the first studied and best examples of a strongly interacting quantum
system that is quite isolated from its environment, is a heavy nucleus.
Experimental data has been amassed for thousands of energy levels but they do not appear to follow
any simple mathematical form, such as the Rydberg formula. Instead it appears that the levels are quite 
random~\cite{shriner1991fluctuation} and can be well characterized statistically.
In 1955, Wigner~\cite{wigner1993characteristic}
%Wigner, E. "Characteristic Vectors of Bordered Matrices with Infinite Dimensions." Ann. of Math. 62, 548-564, 1955.
proposed using random matrices to understand the distribution of energy levels
found experimentally, which turned out to be an extremely deep and insightful approach
to adopt. His reasoning was quite general, and had nothing to do with the precise form of
interactions in the nucleus, as QCD was unknown at the time. Since then, the same random
matrix models have been shown to describe the statistics of a large number of interacting
quantum systems. 

The statistics of the eigenvalues of real symmetric random matrices map on to the classical statistical mechanics of
one dimensional gas with repulsive logarithmic interactions, held together by an external
quadratic potential. The position of a particle corresponds to an eigenvalue,
so at finite temperature, the set of eigenvalues will look like a snapshot
of this thermal system. This leads to a number of interesting properties, and the most clear
signature is to look at a suitably normalized difference between adjacent levels. Because of the
repulsive interactions, this leads to strong energy level repulsion. Wigner showed that the
distribution of energy level differences is very close to
\[
P(s) = \frac{\pi}{2} s e^{-\pi s^2/4}
\]
This fits the data on nuclei quite well~\cite{shriner1991fluctuation}. This also implies no energy
degeneracies.

Because such problems are quite intensive numerically, the first direct calculation
of quantum energy levels showing such a connection occurred in systems with a small number of
degrees of freedom in the semiclassical regime, such as Sinai billiards~\cite{BohigasPhysRevLett.52.1}. The remarkable
similarity of the statistical properties of these energy levels to those of real symmetric random matrices
led to the Bohigas-Giannoni-Schmit conjecture~\cite{bohigas1984spectral}: 
That in fact, any classical system that is strongly
chaotic, has corresponding quantum energy levels that have the same statistics as such random matrices (in the limit
of high energy levels).

\subsubsection{Chaos in classical versus quantum mechanics}

We have seen that there is a well studied path to understanding thermalization in
classical systems, and the key concepts there are chaos and ergodicity. We have also
seen how classically chaotic systems appear to behave when quantized. But it is
still not at all clear how thermalization can occur in quantum mechanics. Here are
some differences between quantum and classical theories:

\begin{enumerate}[i]
\item
There are many extra invariants of motion in quantum mechanics.
For example, any function of the Hamiltonian $f(H)$ because $[f(H),H]=0$. In particular
any projector of an individual energy eigenstates $\pro{E}{E}$ will be a constant of motion. 
These constants of motion are there even with classically chaotic Hamiltonians. This is related
to the next point.
\item 
Another way understanding this is to write the time dependent wave function
as a spectral expansion
\[
\label{eq:psi_t=sum_E_C_E_e_-Et_ket_E}
\ket{\psi(t)} = \sum_E c_E e^{-iEt} \ket{E},
\]
where $c_E = \braket{\psi(0)}{E}$. $\ket{\psi(t)}$ has a constant of motion associated with
every energy eigenvector. In this sense it appears analogous to integrals of motion in
classical mechanics, that give rise to a similar (quasiperiodic) formula for phase space variables.
\item In order to discuss ergodicity, the central concept we used was that of {\em phase space}.
But in quantum mechanics position and momentum do not commute. That is, it is not possible
to simultaneously measure all momentum and positional degrees of freedom. Therefore the whole conceptual
framework illustrated in \Fig{fig:phase_space} will not work. At a semiclassical level, one
can think of wave packets of linear dimensions
$\Delta x$, and momentum dispersion $\Delta p$ chosen such that  $\Delta x\Delta p > \hbar$,
but this wave packet will spread in time and this is only useful in the semiclassical limit.
\end{enumerate}

Aside from these differences in the mathematical structure, at a physical level, 
classical and quantum thermalization appear very different, particularly for a
system with a small number of degrees of freedom.

Consider two classical Sinai billiards in some volume, for example, a box with either hard wall or
periodic boundary conditions.
If we start off the system with arbitrary initial momenta and positions, despite
its small size, we can still say that the system will ``thermalize". \Eq{eq:O_T=int_O_over_int}
will apply in this case and the only dependence that answer will have is on the systems total
energy. Aside from that, there is no dependence on the initial conditions. 
For example, the momentum distribution will be isotropic. This will be the case
for any choice of the radii of the two billiards. However if you make the radii zero, then no
interaction between them is possible, and the system becomes integrable, in which case
average values will depend on the initial conditions, because now the energy of the individual
billiards is conserved. For example, now the momentum distribution will no longer be isotropic.

We can contrast with the quantized version of this problem.
If we start off with two billiards of finite radii and start them off in
an arbitrary initial state,  we can
calculate the time average of an observable. Using \Eq{eq:psi_t=sum_E_C_E_e_-Et_ket_E}, we
can calculate the time averaged expectation value of observable 
$\hat O$~\cite{jancel2013foundations,farquhar1964ergodic,penrose1979foundations},
\[
\label{eq:FineGrainedErgodicTheorem}
\langle \braket{\psi(t)}{{\hat O}|\psi(t)}\rangle_t = \sum_E |c_E|^2 \braket{E}{{\hat O}|E}
\]
where here we assume no degeneracy, as is the case for the above example.
This shows that observables depend on all of the coefficients $c_E$, and therefore
depend sensitively on the initial condition. 

Time averages for ergodic classical mechanical systems only depend on the total 
energy, whereas quantum mechanical systems depend on the details of the initial conditions. 
One might first believe that the reason for this difference is the linearity of quantum
mechanics, and the nonlinearity of classical mechanics.

The perspicacious reader may notice that classical mechanics can be described as a linear equation
quite analogous to the Schroedinger equation~\cite{Koopman1931PNAS17_315K,neumann1932operatorenmethode,von1932zusatze}, 
which evolves a probability amplitude, as opposed to  Liouville's equation, which evolves a probability.
However for these kinds of equations, the eigenvalues are badly behaved~\cite{WilkiePhysRevA.55.27} 
which invalidates the derivation of \Eq{eq:FineGrainedErgodicTheorem} in this case.

At this point, we can see that thermalization cannot occur in general in a quantum system, as we have
shown in the above example how it apparently fails to work for the case of two Sinai billiards.
It went wrong because we have a small number of particles leading to a low density of states. As we shall now see, the situation greatly improves in the large $N$ limit.

%% file: theproblem.tex
\section{Statement of the problem}
\label{sec:theproblem}

The central question of this discussion is to understand why and how an isolated system thermalizes. 

For a large number of number of degrees of freedom, $N$, 
the density of states $n(E)$ is very large. 
This is proportional to $\exp(S(E)$, where $S(E)$ is the entropy. Because usually the entropy
is extensive, $n(E) \propto (\exp(const\times N)$. This means that the average level spacing,
which is inversely proportional to $n(E)$, is incredibly small. This plays an important role in understanding the
nature of thermalization. Unlike classical mechanics, we can see from the discussion in the 
introduction of two billiards, that statistical mechanics is not expected to work for low lying
energy states, and the results obtained will depend on details of the initial state.
But we expect that statistical mechanics will emerge in the large $N$ limit.

To be more precise, we can start by posing a related but easier question. Suppose we have a
pure state $\ket{\psi(t)}$ and an observable $\hat O$. As we discussed in the introduction,
we can consider the time average
of the observable $\langle \braket{\psi(t)}{{\hat O}|\psi(t)}\rangle_t$ which according to
\eqref{eq:FineGrainedErgodicTheorem} can be related to the expectation values of these operators in energy
eigenstates, terms like $\braket{E}{{\hat O}|E}$. 

We want to know why this time average is
in accord with the prescription of statistical mechanics. For an isolated system we can define the
quantum mechanical version of the microcanonical ensemble discussed for classical systems in the
introduction. Because energy levels are quantized, the microcanonical ensemble involves averaging
over all eigenstates within an energy window $\Delta E$. For large $N$, this means that we can take $\Delta E$ to be
very small, still much less than any energy scale in the problem, yet it contains an extremely large number of eigenstates. 
The microcanonical average at energy $E$ becomes: 
\[
\label{eq:DefOfMicroAve}
\mean{\hat O}_{micro,E} = \frac{1}{\cal N}\sum_{E'\in [E,E+\Delta E]} \braket{E'}{{\hat O}|E'}. 
\]
Here $\cal N$ are the number of levels being summed over. What this formula is telling us, is that
to obtain the correct time average of an observable, we calculate the expectation value of $\hat O$
for all energy eigenstates in an energy shell, and then average over all those results.

Why should this microcanonical average be the same as \eqref{eq:FineGrainedErgodicTheorem}? That is the
central question that we want to discuss and is the hardest part of the justification of statistical
mechanics. If that can be established, it is relatively straightforward to understand why the
microcanonical average is equivalent to the canonical one
\[
\label{eq:DefOfCanonAve}
\mean{\hat O}_{canon} = \frac{1}{Z}\sum_E  \exp(-\beta E) \braket{E}{{\hat O}|E}. 
\]
where the normalization $Z$ is the partition function, and $\beta$ is the inverse temperature.
The equivalence of ensembles can be understood by considering a subsystem $A$ much 
smaller than the rest of the system, $B$, and
considering operators $\hat O$ that are local enough to only depend of degrees of freedom in
$A$. Intuitively one can think of $B$ as acting as a heat bath of $A$~\cite{reif2009fundamentals}, 
but more rigorous derivations
using steepest descent can be given~\cite{lax1955relation}. In fact the exponential function can be substituted for many
other functions that decay quickly enough, and these can be shown to be equivalent to the
microcanonical ensemble as well. The extremely fast rise in the density of states, multiplied by
$\exp(-\beta E)$, yields a function that is very strongly peaked at one energy, which is the
essential reason why the result looks microcanonical.

Now let us return to the microcanonical ensemble and discuss how thick an energy shell that we
need to consider.  It is useful to
consider the temperature of a system at energy $E$, assuming that it does thermalize. Mathematically
this means calculating the relationship between average energy and temperature $T$, using the partition function.
The temperature sets a scale for the energy in \Eq{eq:DefOfCanonAve} that tells us qualitatively as the available
energy in single particle excitations. The average energy per particle $(E(T)-E(T=0))/N$, is another scale that
that will be related to the temperature by a function that is independent of $N$ for local Hamiltonians. The change
in the systems physical behavior when an additional energy $k_B T$ is added to it, is negligible. Thus choosing
$\Delta E = k_B T$ in \Eq{eq:DefOfMicroAve}, means that the energy of the states that are contributing, are physically
indistinguishable. Because of the large density of states, the average separation of 
levels is 
\[
\delta E \propto \exp(-S(E)) \propto \exp(-const. N),
\]
and is exceedingly small for macroscopic systems. Within an energy shell of thickness $k_B T$, there are
an exponentially large number of states. Therefore, statistical fluctuations from eigenstate to eigenstate will
be come negligible.

Viewed this way, we can rephrase our central question: why is
\eqref{eq:FineGrainedErgodicTheorem} equivalent to \eqref{eq:DefOfMicroAve}?

%% file: routes.tex
\section{Routes to understanding statistical mechanics}
\label{sec:routes}

Everyone who has taken a course in statistical mechanics has heard
some attempt to justify it. There are a number of different routes
that have been taken and this section discusses the most popular
ones. \footnote{To be clear, many of the papers referenced here
describe these various routes, but the authors of them have often
worked on a number of approaches and would not necessarily disagree
with the criticisms made of these approaches. There are nevertheless strong reasons
to consider each of these approaches.}

\subsection{Typicality}

The idea of typicality is intimately related to empiricism. The Sun has risen
for thousands of years, so we expect it will come up tomorrow. It may not, if say a planet sized
object were to hit the Earth, but barring extremely unlikely events, we expect it to behave
as it always has. Similarly, suppose we look at a ``typical"
quantum state of given total energy, and can show that it almost certainly will obey the laws of statistical mechanics.
Then we expect that in any experiment that we perform on a particular state, it should with very
high probability, obey statistical mechanics. This idea goes back to Schr\"odinger~\cite{schrodinger1989statistical} 
and has led to insights into many questions. Although its purported explanation for thermalization
is criticized here, it is indeed very useful in understanding the
ETH~\cite{tasaki1998quantum,gemmer2003distribution,popescu2006entanglement,goldstein2006canonical,linden2009quantum,muller2015thermalization}.

In order to make this kind of argument more precise, a probability measure needs to be defined on
the space of wave functions, that is, Hilbert spaces.
It is easiest to assume that all wave functions with energy within a small range,
$[E,E +\delta E]$ are weighted with equal probability. When a wavefunction is picked from this
distribution, it defines a system in a pure state. One can ask how well expectation values of
observables agree with the results from a microcanonical distribution.

A wavefunction can be written as in \Eq{eq:psi_t=sum_E_C_E_e_-Et_ket_E}, at $t=0$. There
are an extremely large number of coefficients $c_E$, and typically they will all be nonzero.
The condition on the energy means that to be typical, the $c_E$ are drawn uniformly from 
the surface of a high dimensional hypersphere. 

If we consider $\braket{\psi}{{\hat O}|\psi}$ for any $\ket{\psi}$ with energy within this energy shell,
then its value will depend on the choice of $\ket{\psi}$. We can ask what is 
$\braket{\psi}{{\hat O}|\psi}$ when averaged over such $\ket{\psi}$. We can call this
$\mean{\hat O}_{typical}$. And we can ask how much
$\braket{\psi}{{\hat O}|\psi}$ varies as we change $\ket{\psi}$. 

Because we are averaging over a very large number $c_E$'s, it is not surprising that the
fluctuations in $\braket{\psi}{{\hat O}|\psi}$ are small.
Also $\mean{\hat O}_{typical}$ is almost identical to 
$\mean{\hat O}_{micro}$, as in both cases, an average is being taken over a very
large number of independent wavefunctions. 

Thus if we pick a wavefunction drawn from this uniform distribution,
we would expect $\braket{\psi}{{\hat O}|\psi}$ to be extremely close to the microcanonical
average. This has been shown to be rigorously the case in quite a few respects. 
For any small enough subsystem that is
weakly coupled to the rest of the system, the expectation value of operators in it 
will agree with the canonical average, \Eq{eq:DefOfCanonAve}. In other words, typically small
enough systems look completely thermal and the expectation value of any quantity will agree with
results from the standard result from statistical 
mechanics~\cite{tasaki1998quantum,gemmer2003distribution,popescu2006entanglement,goldstein2006canonical,linden2009quantum,muller2015thermalization}.

The main criticism of this approach is that in real experimental situations, the wave function is {\em not
typical}. If your wave function was typical, then according to the above, you would be in 
thermal equilibrium. If that were the case, all
macroscopic currents, such as those in neuronal action potentials, would be missing. You would not be
able to process any information. Life is incompatible with thermal equilibrium and therefore
we are decidedly {\em not} in a thermal state. 
We are in a state that is far from typical, composed of many macroscopic regions
with different temperatures, chemical potentials, and pressures. The second law of thermodynamics 
also tells us that the Universe is not in equilibrium and therefore not in a typical state. Assuming
that we are overwhelming likely to be in a typical state, we would predict
a universe at constant temperature, and with an entropy that does not change in time. The
fact that you are reading this now, proves that this assumption is incorrect.

But it has been possible to prove many useful theorems about typical states, and these have
found many applications. So typicality remains an important approach to understanding statistical
mechanics.

Another important general criticism of this approach will be given below for why this is not an adequate
explanation of the central problem we wish to understand.

\subsection{Ignorance of the system}

``Statistical" often denotes the use of probabilistic reasoning, which normally connotes
some uncertainty. This has been conflated to mean that statistical mechanics must have a
probabilistic component due to a lack of information~\cite{jaynes1957information,jaynes1957information2,tolman1938principles}. 
Some knowledge is uncertain because we cannot be bothered to find out
more about it, for example the temperature right now in Tulsa Oklahoma. This is an example of
knowledge that is subjective, because another reader could easily look up the answer. On the
other hand, in quantum mechanics, most questions do not have an answer which is certain and this
second class of objective uncertainty has been conflated with the first so
for the moment, let us confine ourselves to classical statistical mechanics where there are
no intrinsic uncertainties in a system's evolution.

There has been a school of thought which has used a lack of knowledge to make physical predictions
about systems at finite temperature. It is indeed true that it is extremely difficult to obtain the
positions and momenta of gas in a room. But if someone did, this would not alter any 
physical outcomes (again at a classical level). Their knowledge wouldn't change the room's temperature, or
pressure. Lack of knowledge at a classical level cannot be used to infer physical quantities.
However if you follow this approach, you can use ignorance to ``derive" the canonical or microcanonical
distributions. 

If one does not know any more than the energy of a system, then one can ask, what is the probability
that it is in a particular microscopic configuration (or state). For example, when tossing a coin,
without any information, you could say {\em a priori}, that there is equal probability that the
outcome is a head or a tail. Similarly if you measured the $z$ component of spin, you could also
assign it equal probabilities. Now suppose you were given 10 weakly interacting spins that were in a magnetic
field. You can then ask what the probability distribution of configurations would be if you were
given the total energy. Using ignorance, you can assign equal weights to all configurations that have the correct
energy. This is mathematically equivalent to a microcanonical ensemble. Despite the fact that
the prediction is based on a lack of knowledge, one can then perform experiments where you
can measure precise spin configurations and see whether they do in fact agree with the
microcanonical prediction. In many cases, of course, they do, giving some confirmation to
this strangely illogical procedure.

This provides an explanation for the laws of statistical mechanics, at least when it works.
The system needs to be in equilibrium and also not integrable, as we will discuss further below.
But it is fundamentally flawed for the reasons above, and incapable of correctly predicting when
the laws of statistical mechanics actually fail. 

There are quite a few mathematical details left out in the brief summary given here, such as the maximization
of the probability distribution, conditioned on energy or particle number constraints. And this involves some 
elegant mathematics. But this does not alter the fundamental flaw in this approach. Even at a classical
level, the correct explanation for statistical mechanics involves ergodicity, which is something
that does not come into this analysis in any way.

The situation is complicated by the addition of quantum mechanics, where the wave function
represents a probability wave, and therefore the system is intrinsically probabilistic. However this
only serves to make the argument more complicated. Fundamentally you can prepare a quantum state and
know a lot about it. For example, in principle, it is possible  to prepare a system so that all of
the $c_E$'s in \Eq{eq:psi_t=sum_E_C_E_e_-Et_ket_E} are known. If this information is added, it will completely
alter the resultant probability distribution according to this sort of reasoning, and will no longer look microcanonical, or
canonical. 

\begin{figure}[]
\begin{center}
\includegraphics[width=1\hsize]{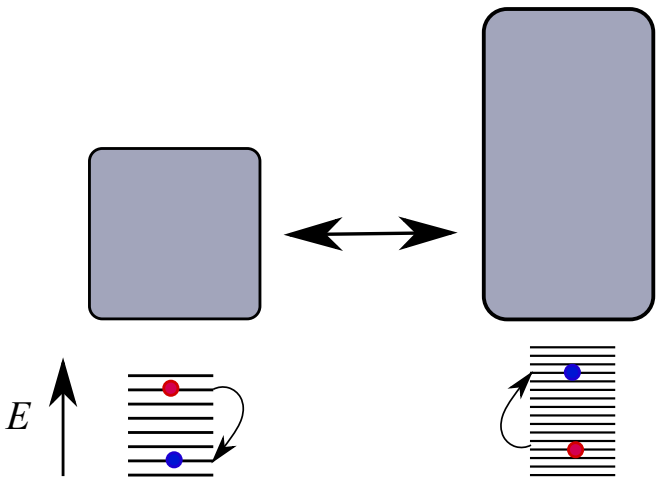}
\caption
{ 
A small and system coupled weakly to each other. For short times, each system can be thought of as
being isolated. Initially the energy levels of each of the systems is shown by the red dots. After
a long time, they transition to new levels, shown as blue dots. After a long time, the smaller
system explores a large number of separate states.
}
\label{fig:open_systems}
\end{center}
\end{figure}

\subsection{Open systems}

With open systems, the system of interest is contact with a much
larger one, considered to be a ``heat bath".  The bath has properties
that make it straightforward to obtain statistical mechanics for
the system of interest. One justification for this is that it is
impossible to have a truly isolated system, and therefore one must
always consider a smaller system that is connected to a much larger
one~\cite{blatt1959alternative,gemmer2004quantum}. The smaller
system exchanges energy with the larger system, and it is possible
through a number of arguments, to derive the canonical distribution
for the smaller system.

One intuitive explanation assumes a very weak coupling between the
two systems as shown in Fig. \ref{fig:open_systems}. For times that
are short enough, the two parts of the system will appear isolated,
but for longer times, there will be an exchange of energy between
them. Thus the energy in the smaller system will increase and
decrease, sampling states of different energies. A time average
over the smaller system will then yield a weighted average over
different energy levels, depending on how much time is spent in the
different states. It is straightforward to argue that this will be
highly peaked function of energy, and thus equivalent to an ensemble
average, for example \Eq{eq:DefOfCanonAve}. This kind of argument is often discussed in 
introductory textbooks, because of its relative simplicity~\cite{reif2009fundamentals}. Not only does this
approach sidestep the need for understanding the intricacies of a
closed system's dynamics, but can be very useful to derive approximate
equations for the smaller system's evolution, and these can be very
valuable~\cite{lindblad1976generators,lindblad2001non}.

But what if the complete system has a quadratic Hamiltonian? This would not thermalize, so clearly
there is a problem with the above approach. But we could
allow the complete system to weakly couple to an even larger system, and hope that this would save this approach,
unless the larger system was also quadratic, which would mean coupling yet again.
These turtles all the way down arguments have an intuitive appeal but lack logical clarity.

\subsection{Litmus test for these explanations}
\label{subsec:LitmusTest}

If we refer back to Fig. \ref{fig:phonon_relaxation}, and its discussion in the introduction, we
see that not all system thermalize. If an explanation cannot distinguish between integrable and 
generic {\em non-integrable} dynamics, and predicts all such systems thermalize, it has to be incorrect. The above
explanations all are incapable of making this distinction. At a classical level, explanations,
such as using ignorance, cannot be correct as was already noted. 

The fact that most systems
do obey statistical mechanics, does not make these explanations correct, but  it does not make
these ideas without merit. Typicality is still a very useful 
mathematical result, that has applications to many areas even including numerical simulations. 
Reasoning using ignorance,
or open systems, are relatively simple and intuitive. They
provide a lot of intuition and guidance in understanding physics problems. When a system does
thermalize, it is often very useful to divide it up into smaller and larger pieces, where open system
arguments can be used to understand statistical fluctuations. Another interesting approach uses
the large number of coefficients in  \Eq{eq:FineGrainedErgodicTheorem} expected in
experimentally realistic conditions to find a agreement with statistical mechanics~\cite{reimann2008foundation}.
Also it would be remiss not to mention an insightful approach provided by Von Neumann's ``Quantum
Ergodic Theorem"~\cite{von2010proof},
which was quite recently developed further~\cite{goldstein2010normal,goldstein2010long}. As shown recently, the
assumptions that go into this theorem are essentially the same as assuming the ETH~\cite{rigol2012alternatives}.

A more predictive and fundamental understanding of thermalization, needs to be able to distinguish
between the different physical behavior of generic systems, which are chaotic, and integrable systems. 
The analog of chaos in quantum mechanics has to be better understood in order to make progress
understanding the origin of statistical mechanics. 

The route taken here is to try to parallel the discussion of classical systems of the
introduction. We want to understand if there is some analog of ergodicity and chaos that can be used
to understand quantum statistical mechanics, that is \Eq{eq:DefOfMicroAve}. What special features
of generic quantum systems allow thermalization, where as integrable systems do not?

%% file: randommatrix.tex
\section{Relation to Random Matrices}
\label{sec:RelRandMatrices}

As was discussed in the introduction, there appears to be an intimate connection between strongly
interacting quantum mechanical systems, and random matrices. The micro-structure of the energy levels
appears identical for systems with high density of states. Other important physical properties also
appear to be intimately tied to random matrices as well. In particular, how do we expect energy eigenstates
of quantum mechanical systems to behave, and how is this related to random matrices?

\subsection{Semiclassical Limit}
\label{subsec:SemiclassicalLimit}

Studies of the semi-classical limit, of hard sphere systems, such as those shown in Fig. \ref{fig:SinaiBunimovich} 
have largely answered how energy eigenstates behave in the semiclassical limit.
It's simpler conceptually to think of a single particle in $3N$ dimensional space, with a momentum
$\bf p$ and so that the Bunimovich stadium and Sinai Billiards
just correspond to different boundary conditions of a stadium problem in higher dimension. 

The energy eigenstates of non-interacting particles in free space
are plane waves, in other words, can be chosen to be in a definite
momentum state $\ket{\bf p}$.  Classically, at a given energy, a
total momentum momentum is $p^2/2m$.  When go over to quantum
mechanics, the general solution will be the sum (or integral) over
plane waves on the energy shell, with amplitudes chosen to match
the boundary conditions, in other words we are integrating over
plane waves going at different angles.  The analogy of classical
ergodicity in this case, would be to have the amplitudes of the
plane waves uniformly distributed in angles, rather than focused
in particular directions.  A number of theorems have been proved
that show that this is the case \footnote{Note that for the Bunimovich
stadium it can also be proved that there exist some subset, that
becomes vanishingly small in the high energy limit, that are not
ergodic~\cite{hassell2010ergodic}.} in the semiclassical
limit~\cite{ShnirelmanErgodicBilliards,shnirelman1993addendum,zelditch1987uniform,colin1985ergodicite,helffer1987ergodicite,berry1987bakerian,Berry0305-4470-10-12-016}.
There is further numerical evidence supporting the idea that the
amplitudes of these momentum states behave
randomly~\cite{aurich1999maximum}.

The scope of the validity of the microcanonical average in the limit as $\hbar \rightarrow 0$, was further
expanded by work that considered other Hamiltonians in the semiclassical limit and showed reasonable agreement
with the microcanonical ensemble even for systems with a few degrees of freedom~\cite{FEINGOLD1985344,FeingoldPhysRevA.34.591}.

\begin{figure}[]
\begin{center}
\includegraphics[width=1\hsize]{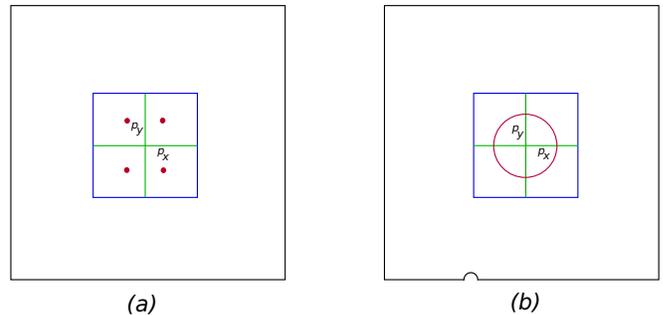}
\caption
{ 
A comparison of a semiclassical integrable and non-integrable system comprising a particle moving in a stadium.
The black stadiums in (a) and (b) represent the hard walls, and superposed in the middle are probability densities of 
energy eigenstates (red) shown in the momentum representation, characterized by the momentum $p_x, p_y$.
(a) Here the classical motion is integrable and only four momenta have nonzero probabilities. Classically
corresponding to this, 
the particle will bounce off the walls alternating between these four discrete values of momenta.
(b) The same situation where a small change to the boundary has been made, making the system non-integrable.
Classically the momentum distribution will spread out isotropically over a circle.
In the semi-classical case, the probabilities will also be spread out but their amplitudes will be uniformly
distributed but random on this circle.
}
\label{int_non_int_comparison}
\end{center}
\end{figure}

Therefore we have fairly good picture of a typical semiclassical energy
eigenstate of hard sphere systems, that is eigenvalues of the Laplacian
operator with appropriate boundary conditions. It will look like a random superposition of
plane waves uniformly distributed over an energy shell. This is illustrated in Fig. \ref{int_non_int_comparison}
for the integrable case of a particle in a  rectangular box in two dimensions, and rectangular box that has
a small circular protrusion.

The randomness in the wavefunction is also mirrored by the randomness that appear
in the distribution of eigenvalues which follow random matrix theory as well mentioned
in the introduction\cite{bohigas1984spectral}. The eigenvectors of a real symmetric random matrix
will form a set of orthogonal random vectors, very similar to the orthogonal set of random
vectors seen in the momentum representation of semiclassical billiards.

On the other hand, an integrable system will have completely different properties in the semiclassical
limit, and random matrix models do not apply. Rather an extra set of invariants (such as the phonon example
of the introduction) are used to construct energy eigenstates.

An important question to ask is whether or not the random matrix analogy only works
in the semiclassical limit of small $\hbar$, or if it can be extended to systems
where $\hbar$ cannot be taken as small. Underlying the analysis of energy level statistics and the randomness of wavefunctions
was the need for a high density of energy levels. For small $N$, this would require a semiclassical approach, but
for large $N$, systems at low temperatures and their associated energies, have correspondingly high
density of states. We will ask in general how we can understand energy eigenstates for finite $\hbar$.

\subsection{Perturbing an integrable model}
\label{subsec:PertIntModel}

If a system is integrable, it will never thermalize. If now turn on an interaction to break integrability, 
we can ask what will happen. As discussed in the introduction in the context of KAM theory,  
for large $N$, it is generally believed that only 
a very small interaction is needed to destroy orbits with regular motion. And with Sinai Billiards, any
non zero billiard radius will immediately destroy integrability. 
With the large $N$ classical situation in mind, in a quantum treatment, can we understand 
how a system will transition between integrable and non-integrable behavior? We will try to follow the same path as
for the semiclassical limit of hard sphere systems~\cite{deutsch1991quantum}.

Suppose we consider the Hamiltonian $H_0$ of the integrable system. We now add in a very weak integrability breaking
term $H_1$ of order $\epsilon$, 
\[
H = H_0 + H_1.
\]
$H_1$ could be a two body interaction between particles in a gas, for example. As we have seen for hard core systems,
semiclassically, in the basis of $H_0$, the eigenstates appear to be random superpositions. And the behavior 
of energy level spacings that we discussed, suggest that in this integrable basis, $\braket{E^0_i}{H_1|E^0_j}\equiv h_{ij}$ can be
can be taken to be a symmetric random matrix, but with the elements $h_{ij}$ being small.

Although there is some justification for this choice, it is still not clear how the elements $h_{ij}$ should be chosen.
If we couple states of vastly different energies together, it will lead to catastrophic effects on the dynamics. 
The eigenvectors will now become completely delocalized in the energy basis of $H_0$, yielding nonsensical results.

Therefore from a physical
perspective, $\braket{E^0_i}{H_1|E^0_j}$ cannot have statistics that are independent of $i$ and $j$. In fact, if one calculates to
second order in perturbation theory, the size of these elements, phase space arguments~\cite{deutsch1991quantum} show that 
$h_{ij} \sim \exp(-\beta |E_j-E_i|)$ for $\beta |E_j-E_i| \gg 1$. Here $\beta$ is the inverse temperature corresponding to a
system with average energy $E_i$ (or $E_j$ because the difference between the two is so small in this context). 
With matrix elements decreasing away from the diagonal, we can take $h_{ij}$ to be a banded symmetric random 
matrix. Numerical confirmation of this bandedness 
had been given in the semiclassical limit quite early~\cite{feingold1989semiclassical}.
A more recent study of Hubbard models~\cite{genway2012thermalization} showed a general
banded structure for many quantities of interest. Bandedness is a 
crucial component of this matrix, as without it, the expectation values of operators would be
completely unphysical, dominated by the highest energy states of the system. Further evidence
for the banded nature will be given when discussion the ETH in the next section.

Thus the model in the integrable energy basis looks like a diagonal matrix, with increasing diagonal elements,
$\braket{E^0_i}{H_0|E^0_i}$, to which a banded symmetric random matrix is added~\cite{deutsch1991quantum}. 
We need to understand some basic properties of eigenvectors of this kind of matrix. 
If the eigenvectors become still become delocalized
this model would behave incorrectly. Fortunately, for such a model, or ones quite similar~\cite{wigner1957BandedRandomMatrices}, 
it can be shown that the eigenvectors are localized in energy. 

This localization of eigenvectors in the integrable energy basis
is an extension of what is found for billiard systems semiclassically.
It says that a typical eigenstate of a non-integrable system is the random superposition of integrable states in 
some narrow energy shell
\[
\label{eq:E_i=sum_c_ij_E_j}
\ket{E_i} = \sum_j c_{ij}\ket{E^0_j}
\]
and the $c$'s are the matrix of eigenvectors $\braket{E_i}{E^0_j}$.  Because $H_1$ is random the $c_{ij}$ will be also.
Averaging over different random realizations of $H_1$, $\mean{c_{ij}^2}_r$ is strongly localized around $i = j$. 
This equation suggests that we can think about the energy eigenstates of the non-integrable system 
as the random superposition of the integrable eigenstates. Let's compare this again to the sudden
transition from integrable to non-integrable behavior you see with Sinai Billiards, going from a radius $r=0$ to $r>0$.
For $r=0$, you have an ideal gas which will never thermalize. But $r > 0$ but very small, eventually the system will
thermalize due to particle collisions. In the quantum case, with nonzero $H_1$, you will get \Eq{eq:E_i=sum_c_ij_E_j}.
How small does $H_1$ have to be? It turns out to be very small for large $N$. The relevant energy scale here is the energy level 
separation which decreases exponentially with $N$, so for an Avogadro's number size system, this will involve numbers of 
size $\exp(-10^{23})$. We will discuss later, the transition for much smaller numbers of particles, as you would get in a numerical
simulation or a cold atom experiment.

\subsection{Expectation values of nonintegrable eigenstates}
\label{subsec:ExpNonintegrableEigenstates}

The right hand side of \Eq{eq:FineGrainedErgodicTheorem} involves the expectation values of an operator $\hat O$ in energy
eigenstates, $\braket{E}{{\hat O}|E}$. Therefore this quantity is central to calculating time averages of operators, which in 
turn should be the same as microcanonical averages if statistical mechanics is to hold.
\Eq{eq:E_i=sum_c_ij_E_j} suggests that we can calculate such expectation values in terms of the energy eigenstates of
integrable systems. Because the statistical mechanics of $H$ and $H_0$ should be the same to $O(\epsilon)$, we will see that
expressing averages in terms of eigenstates of $H_0$ will be quite informative.

We can write 
\[
\label{eq:<EiOEi>=sum_cc<>}
\braket{E_i}{{\hat O}|E_i} = \sum_{kl} c_{ik}c_{il}\braket{E^0_k}{\Ohat | E^0_l}
\]
But this involves the random eigenvectors $c$. These have mean zero and their statistics
can be calculated reasonably well. One can ask what is the value of $\braket{E_i}{{\hat O}|E_i}$
averaged over some small energy window of $E_i$, and what is the fluctuation in  $\braket{E_i}{{\hat O}|E_i}$.
This is equivalent to averaging over the $H_1$ $\langle \dots\rangle_r$.
Without going through the technical details, it is not much of a surprise to find that the cross terms in 
\Eq{eq:<EiOEi>=sum_cc<>} vanish, yielding~\cite{deutsch1991quantum}
\[
\label{eq:<EiOEi>=sum_csq<>}
\braket{E_i}{{\hat O}|E_i} = \sum_{k}  \langle c^2_{ik}\rangle_r \braket{E^0_k}{\Ohat|E^0_k}
\]

We can also consider the fluctuations in this expectation value 
\[
\sigma^2_i \equiv Var(\braket{E_i}{{\hat O}|E_i}) 
\]
and show~\cite{deutsch1991quantum} it is extremely
small proportional to $\exp(-S(E)) \propto \exp(-const\times N)$. 
Because the $c_{ik}$'s are sharply peaked around $E_i = E_k$, \Eq{eq:<EiOEi>=sum_cc<>} gives 
the {\em microcanonical average} of $\Ohat$.

The intuitive picture of how non-integrable energy eigenfunctions appear from an integrable model, is not unlike
what happens in the semiclassical case of hard walls, as illustrated in Fig.  \ref{int_non_int_comparison}.
An energy eigenstate is formed by the random superposition of states with very similar energies. Because of the 
amplitudes of all of these states are random, when forming expectation values of an operator, only its
diagonal matrix elements contribute. This then gives expectation values of the operator averaged over
an energy shell, which is precisely the microcanonical average.

It is also interesting that this arguments if we start an non-integrable point and perturb it which suggest
that it may be of considerable generality.

\subsection{Off diagonal elements}

It is also possible to calculate the properties of off-diagonal matrix elements $O_{ij} \equiv \braket{E_i}{{\hat O}|E_j}$
in this model~\cite{reimann2015eigenstate}. Their mean is zero and their 
variance can also be shown to be of magnitude to 
$\sigma^2_i \propto \exp(-S(E)) \propto \exp(-const\times N)$. Their value will go to zero as $|E_i - E_j|$ become large.  
These off diagonal elements are important in determining
the dynamical correlations, rather than expectation values of observables, averaged over time.

%% file: eth.tex
\section{The Eigenstate Thermalization Hypothesis}
\label{sec:ETH}

In the last section shows that for fairly general reasons, we expect that $\braket{E}{\Ohat|E}$ fluctuates
very little as $E$ is varied and gives results in accord with the microcanonical
distribution~\cite{deutsch1991quantum}. This leads us to the ``Eigenstate Thermalization Hypothesis" (ETH). 

\subsection{Statement of the Hypothesis I}
\label{subsec:StatementOfETHI}

The term ``Eigenstate Thermalization"
appears to have been first coined by Mark Srednicki~\cite{srednicki1994chaos} as a succinct
description of how a single eigenstate can be thought of, as being in an equilibrium thermal state, 
in the sense now described. 

Consider a finite isolated system, with a non-integrable Hamiltonian with $N$ degrees of freedom. The
eigenstates $\ket{E_i}$ are solutions to  $\Ham \ket{E_i} = E_i \ket{E_i}$. The solutions should
also be separated by symmetry sector~\cite{santos2010onset}. For example, as mentioned in the introduction, total momentum
is often conserved, particularly in homogeneous systems with periodic boundary conditions. Only in a single
sector can we assume that the $E_i$'s are non-degenerate.
\begin{conjecture}
For a large class of operators, $\Ohat$, we consider its expectation values $O_{ii} \equiv \braket{E_i}{\Ohat|E_i}$
as a function of $i$. We also consider the microcanonical average $\mean{\Ohat}_{micro,E_i}$ as defined in
\Eq{eq:DefOfMicroAve}. Then
\[
\label{eq:Oii=mean+delta}
O_{ii} = \mean{\Ohat}_{micro,E_i} + \Delta_i
\]
where $\Delta_i$ has zero mean and $\Delta_i^2$ has a magnitude of order $\mean{\Ohat^2}_{micro,E_i}\exp(-S(E))
\propto \mean{\Ohat^2}_{micro,E_i}\exp(-const.\times N)$.
\end{conjecture}

The sole fact that $O_{ii}$ varies very little among neighboring eigenstates
implies that $\mean{O_{ii}}$ averaged over a small energy window must
give the microcanonical average. Thus the ETH is really a statement about the small size of fluctuations
of expectation values between eigenstates. It says that the microcanonical average for non-integrable
systems, for most purposes, does not need to be taken at all, and a single eigenstate can be used.

The most unclear part of this statement is the class of operators to which it applies. There are clear
examples of where this fails: For any function $f$, $\Ohat = f(\Ham)$ commutes with $\Ham$ and therefore
$\Ohat_{ii} = f(E_i)$. If $f$ is sufficiently poorly behaved, for example a $\delta$ function, then this
will violate the hypothesis.

But such operators are global, involving all of the degrees of freedom in a system. The ETH is believed to work well
for operators only involving a few degrees of freedom; for example, operators involving the momenta of three particles,
and these do not have to be in the same region of space. 
But the exact limits of where it works and where it fails are still not clear. It is generally believed to be valid
in most non-integrable systems when $\Ohat$ involves operators in some local region. 
In fact it has been argued that for entanglement entropy in the case of two weakly coupled systems similar to the
random matrix models considered above, that the smaller system can be almost 
as large as half the system~\cite{deutsch2010thermodynamic}. And this bound has been argued to hold for more
general quantities for non-integrable systems as well~\cite{garrison2015does}.

The exceptions to the ETH are for systems obeying 
Many Body Localization~\cite{AlthshulerPhysRevLett_78_2803,BaskO20061126,imbrie2016many,HusePhysRevB_88_014206,nandkishore2015many}. 
These are very unusual systems
that violate the ETH even though the system is non-integrable.
We will discuss the validity of the ETH further in 
subsequent sections. 
%but for now we will turn to the more pressing question, of how this explains the process of thermalization.

The way that the ETH has been defined above, is still a bit sloppy in another way. When we gave the size of the $\Delta_i$'s, 
it is
exponentially small in $N$. However Hilbert space is exponentially large and we have not defined the distribution
of $\Delta$'s. It still might be possible that an exponentially small fraction of eigenstate do not obey the ETH and have
expectation values significantly different from the microcanonical ensemble. This can 
be thought of as the {\em weak} ETH\cite{BiroliPhysRevLett.105.250401}.

On the other hand, if we are to say that $O_{ii}$ is always very close microcanonical, for all $i$, this 
means we have the {\em strong}
ETH. One might think that the distinction between the weak and strong ETH is not important. However even
for integrable systems, typical states as noted above, give thermal averages. In fact, all but a vanishingly
small number of eigenstates, will be in this category and can exhibit the weak ETH~\cite{muller2015thermalization}. 
There have been some interesting non-integrable models that have been devised, 
that show an exponentially small number of microcanonical violating
eigenstates~\cite{mori2016weak,mori2017thermalization}, but their interpretation is still the
subject of some debate~\cite{mondaini2017comment,shiraishi2017reply}.

\subsection{Statement of the Hypothesis II}
\label{subsec:StatementOfETHII}

The statement of the ETH has been further expanded to include the behavior of off-diagonal matrix elements.
This is not necessary to understand how time averages are equivalent to ensemble averages. However, among
other things, it is important in the understanding of dynamic correlation functions, and the approach to 
equilibrium~\cite{srednicki1994chaos,srednicki1999approach,KhatamiPhysRevLett.111.050403,DAlessio2016quantum}. 
Here we more generally consider $O_{ij} \equiv \braket{E_i}{\Ohat|E_j}$. For $i\ne j$ the hypothesis is that,
$O_{ij} = \Delta_{ij}$ where $\Delta_{ij}$ appears stochastic and 
has similar statistics to those described for \Eq{eq:Oii=mean+delta}, 
but now there are two relevant energies $E_i$ and $E_j$. 
These elements are therefore extremely small. Because the statistics of this quantity should vary
smoothly as a function of $E_i$ and $E_j$, we can try to quantify this for a single system
by performing averages over the nearby energy levels of $i$ and $j$, $\mean{\dots}_n$,
so that we can write 
\begin{conjecture}
\[
\mean{|O_{ij}|^2}_n  = \mean{{\Delta_i}^2}_n F(E_i,E_j)
\]
with $\Delta_i$ is the fluctuation in the $O_{ii}$'s described in \Eq{eq:Oii=mean+delta}.
$F$ is a function of  order unity for $E_i=E_j$ that goes to zero as $|E_i-E_j|$ becomes large.
\end{conjecture}

%\mean{\O_{ij}^2}_n  = \mean{\Ohat^2}_{micro,\ov{E}}\exp(-S(\ov{E})) f(\ov{E},E_i-E_j)$

Defining $\ov{E}_{ij} \equiv (E_i+E_j)/2$, this can be written somewhat more symmetrically as

\[
\label{eq:Oij=Delta}
\mean{|O_{ij}|^2}_n = \mean{\Ohat^2}_{micro,\mean{E}_{ij}} f(\ov{E}_{ij},E_i-E_j)\exp(-S(\ov{E})/2)
\]
where now $f(\ov{E}_{ij}, E_i-E_j) = F(E_i,E_j)$, describes a function that goes to zero
when $|E_i-E_j|$ becomes large.

The $O_{ij}$ cannot in general be zero however.
For example if we consider an operator $\hat A$ being the square root of $\hat O$, that 
is ${\hat O} = {\hat A}^2$, then 
\[
\braket{E_i|({\hat A} - A_{ii})^2}{E_i} = \sum_{j\ne i} |A_{ij}|^2
\]
The left hand side is not small and is non-negative. The right hand side are terms that appear
in the off-diagonal matrix elements that appear in this hypothesis.
What this implies is that the off-diagonal matrix elements must be quite evenly distributed
or this hypothesis would be violated.

\subsection{Thermalization}
\label{subsec:thermalization}

Now we return to the situation illustrated by \Fig{fig:phonon_relaxation}. An isolated system is put in a nonequilibrium state,
and we ask if it will eventually return to equilibrium or stay in a nonequilibrium state. To identify if the system 
is in equilibrium, we ask if time averages of observations are in agreement with equilibrium statistical mechanics.
This requires that $\langle \braket{\psi(t)}{{\hat O}|\psi(t)}\rangle_t$ gives the average predicted from statistical mechanics.

It is worth noting that the averages predicted by quantum statistical mechanical ensembles are incorrect in situations
where the system is in a macroscopic superposition, as exemplified by Schroedinger's cat. We should expect that a precise
treatment of time averages will yield a result that can include this possible initial state.

\Eq{eq:FineGrainedErgodicTheorem} tells us how time averages can be related to expectation values of eigenstates.
If we assume the ETH, then this becomes to within small corrections
\[
\label{eq:TimeAveWithETH}
\langle \braket{\psi(t)}{{\hat O}|\psi(t)}\rangle_t = \sum_E |c_E|^2 \mean{\Ohat}_{micro,E_i}
\]
If the coefficients $c_E$ are strongly clustered around one energy, then because $\mean{\Ohat}_{micro,E_i}$ is smoothly
varying, this will also yield the microcanonical, or equivalently, the canonical average. On the other hand, it the $c$'s are
not clustered around one energy, as is the case for a macroscopic superposition, then this time average 
will yield microcanonical values of observables weighted over the probability that they are in one of those superpositions.
This is precisely what one would expect should happen in such cases.

Therefore for long enough times, the system will eventually return to a state
of thermal equilibrium, at least when probed with observables for which the ETH is satisfied.
This argument does not tell us how long one has to wait. The times necessary for 
\Eq{eq:FineGrainedErgodicTheorem} to be satisfied can be extremely long. But at the same
time we also know that relaxation for some systems are extremely long even
for an open system in contact with a heat bath. All we can say from the ETH is that equilibrium
will eventually be achieved. 

The crucial qualitative idea behind the ETH, is that each eigenstate is itself ``thermal", giving the same results as for
averages in open systems in contact with a heat bath. One way to think about this is that a small number of
degrees of freedom of the isolated system can be considered a subsystem and the rest of the system can act as a heat bath
which is expected to yield thermal properties for the degrees of freedom being observed.

\subsection{Does thermalization imply the ETH?}

It is worth considering if there may be some alternatives to the ETH 
which can also explain the experimental observation that all real systems
thermalize. It is also not clear whether or not the ETH is really a useful way of understanding thermalization
because, as stated above, even integrable models satisfy the {\em weak} ETH. 

From a theoretical perspective, both of these questions have been addressed. Suppose that we have an isolated
system and view it as being divided into two, a system of interest, $A$ and the rest, $B$, which can be thought of as a bath
for $A$. We now consider starting the system out of equilibrium, in an arbitrary
product state of $A$ and $B$. If all such initial states result in the thermalization of $A$, this requires that the ETH
hold. But it requires that the ETH hold in the {\em strong} sense, that all energy eigenstates are thermal. If this was not
the case, there would be certain initial product states that would not thermalize. 

Therefore what we have learned is that most systems have eigenstates 
that give microcanonical answers, whether or not they thermalize properly or not. In order to get thermalization,
we need all eigenstates to give microcanonical averages.

This situation is not completely settled however, because it is not clear that all initial product
states are experimentally realizable. And there are certainly examples of systems that do not obey this {\em strong} version
of the ETH, but rather the {\em weak} version, but will still be able to thermalize some large class of initial states, but certainly
not all of them~\cite{mori2017thermalization}. It would indeed be very interesting if there existed 
some experimental isolated systems that could be shown to fail to thermalize from certain
states that were carefully prepared. This would imply that statistical mechanics
was not generally valid, opening up many interesting possibilities. Barring this possibility, this is quite compelling
evidence that the ubiquity of thermalization seen experimentally, implies the strong version of the ETH.

\subsection{Entropy and the ETH}
\label{subsec:EntropyandETH}

Another important quantity that is still not well understood is the entropy of a system and indeed
has many definitions.
The Von Neumann entropy for a system with density matrix $\R$ is $S_{VN}(\R) = - \tr \R\ln\R$.
For an isolated system in a pure state this is zero.
The von Neumann entropy represents a lack of knowledge about a system, but not what is measured in thermodynamic
experiments. But is there a way of defining thermodynamic entropy in an isolated system?

The entanglement entropy is often used as a measure of mutual information between two systems. Suppose
we take an isolated system in a pure state, and divide into two subsystems $A$ and $B$. Then the reduced
density matrix for $A$, $\R_A = \tr_B \R$, will in general become mixed because it is entangled with $B$.
The entanglement entropy between $A$ and $B$ is defined to be $S_{AB} \equiv S_{VN}(\R_A)$.

We cannot directly apply the ETH to this problem because the entanglement entropy cannot be written as the
expectation value of an observable. But the mathematics of this problem are quite closely related to the ETH. 
Theoretical arguments\cite{deutsch2010thermodynamic} and numerical work\cite{deutsch2013microscopic,santos2011entropy}
indicate that this is indeed related to the thermodynamic entropy. For homogeneous systems
in the limit of large $N$, finite energy
eigenstates give an entanglement entropy that is equal to the thermodynamic entropy of the smaller of $A$ and $B$. 
This gives an explicit prescription for how to relate thermodynamic entropy to the microscopic
description of the system in terms of its wavefunction. For integrable systems, this connection no longer
holds for energy eigenstates, which shows the underlying importance of quantum chaos in
the validity of thermodynamics. 

This way of describing entropy, via entanglement, has no clear classical analog. Classically,
entropy can be thought of as the amount of phase space explored by the system. To understand
how this can increase in time, degrees of freedom are often coarse grained. 
It has been recently been shown~\cite{ObservationalEntropySafranekEtAl2017,SafranekDeutschAguirreObservationalEntropyLongPaper} that this idea of coarse graining can be extended to quantum mechanics
by making sequential observations of different observables. For example, one can observe coarse grained
positional degrees of freedom and then energy. This allows one to calculate the probabilities
of each of these coarse grained bins, and  construct a Gibbs entropy. For non-integrable systems
by using the same random matrix model employed in understanding the ETH~\cite{deutsch1991quantum}, 
this can also be shown to lead also to the thermodynamic entropy when the system is in an energy
eigenstate.

With either definition of entropy, one ends up with the same thermodynamic entropy as one would have
starting in a thermal state. But it is necessary for the system to be non-integrable in order for these
results to hold.

%% file: numerics.tex
\section{Numerics} 
\label{sec:numerical}

Pioneering initial numerical work on quantum systems with a large number of degrees of freedom~\cite{jensen1985statistical} 
was hampered by
the limited computational power available at the time. This led to a slightly unclear picture of
eigenstate thermalization. At these system sizes, there was not a large distinction present 
between integrable and non-integrable systems,
but the agreement with statistical mechanics was seen to depend on the choice of the observable
and ``good" ones appeared to be necessary in order to obtain this agreement. With larger system
sizes that are easily achievable on todays computers, the distinction between integrable and
non-integrable systems is much more apparent, and it is also clear that the agreement with
statistical mechanics holds over a much wider class of observables, in agreement with the above
theoretical arguments. 

The main tool for studying the ETH numerically is exact diagonalization. This is a technique that
is used to diagonalize the Hamiltonian of a discrete system. For example, Hubbard models, that
allow hopping of particles between different sites with some local interaction between
particles. Another commonly used type of system are ones involving spin degrees of freedom on a lattice.
Because of the exponential growth of Hilbert space with the number of particles, and lattice sites, 
only relatively small systems are accessible this way, in the neighborhood of 25 lattice sites.
There have been a wide range of studies of this type that have given us enormous insight into
the nature of thermalization. The ETH has been observed in a wide variety of these lattice 
systems~\cite{rigol2008thermalization,santos2010onset,RigolPhysRevLett.103.100403,RigolPhysRevA.80.053607,
              RigolSantosPhysRevA.82.011604,KhatamiPhysRevLett.111.050403,PhysRevA.90.033606,
              biroli2010effect,NeuenhahnPhysRevE.85.060101,SteinigewegPhysRevE.87.012118,
              BeugelingPhysRevE.89.042112,KimPhysRevE.90.052105,SteinigewegPhysRevLett.112.130403,
              KhodjaPhysRevE.91.012120,BeugelingPhysRevE.91.012144,FratusPhysRevE.92.040103,geraedts2016many}.

Today, even on a modest laptop, useful information can quite easily be obtained. As an example, consider 
for hard core bosons(HCB) in one dimension, with nearest neighbor (NN) and next nearest neighbor
(NNN) interactions that cannot occupy the same site~\cite{santos2010onset}. They evolve according to the Hamiltonian
\[
\label{boseHam}
\begin{split}
H_B =  \sum_{i=1}^{L} &\left[ -t \left( b_i^{\dagger} b_{i+1} + h.c. \right) +V n_i   n_{i+1} \right.\\
&- \left. t' \left( b_i^{\dagger} b_{i+2} + h.c. \right) +V' n_{i}  n_{i+2} \right].
\end{split}
\]
Here we are summing over all lattice sites $L$.  $b_i$ and $b_i^{\dagger}$
are the boson annihilation and creation
operators, respectively, for site $i$. $n_i = b_i^{\dagger} b_i$ is the boson local density operator.
The NN and NNN hopping 
strengths are respectively $t$  and $t'$. The interaction strengths are $V$ and $V'$ respectively. 
Here we take $\hbar = t =V=1$.

This model is considered with periodic boundary conditions, meaning that there is translational
invariance and particle number conservation. Therefore $H_B$ divides into independent blocks
corresponding to different total momenta $k$. In this example we take $k=1$.

\begin{figure}[]
\begin{center}
\includegraphics[width=1\hsize]{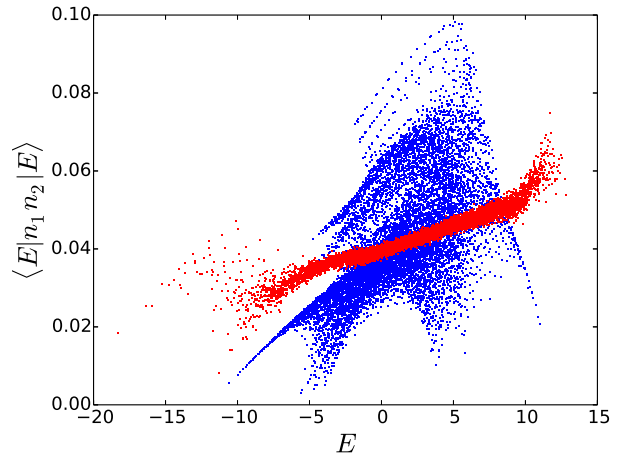}
\caption
{ 
Numerical results for the hard core boson model of \Eq{boseHam} for 6 particles on 17 lattice sites.
The expectation value for the observable $n_1 n_2$, corresponding to the probability that two
particles will be next to each other, plotted for different energy eigenstates. 
The blue points represent the integrable case $V' = t' = 0$,
whereas the red points correspond to the non-integrable case $V' = t' = 0.96$
}
\label{fig:n2}
\end{center}
\end{figure}

We can compare a non-integrable with an integrable choice of parameters to see how expectation
values depend on energy. The choice
$V' = t' = 0.96$ corresponds to a non-integrable set of parameters, while $V' = t' = 0$ is integrable. 
Here, somewhat arbitrarily, we consider the expectation value of $n_1 n_2$, in different energy eigenstates,
with $L=17$ lattice sites and the number of particles of $6$. \Fig{fig:n2} plots these expectation values for these two cases.

In general, as the size of a non-integrable system increases, the fluctuation in expectation
values decreases. Unfortunately, due to the exponential growth of Hilbert space,
one does not expect to be able to simulate systems of even $100$ particles in the future, except
perhaps with quantum computers. However much has been
learnt about such isolated quantum systems by performing such numerical experiments, partly because
one expects a rather rapid decrease in fluctuation as predicted by the \Eq{eq:Oii=mean+delta}, and verified
numerically~\cite{beugeling2014finite}.

Many interesting things have been learned from studies of this kind, and allow us to answer questions that are very
difficult to study by purely analytic means.

For finite size simulations, one can tune how far one is away from  integrability. For example, in the hard core 
boson model above, those parameters are $t'$ and $V'$. For the small systems that are accessible by exact
diagonalization, the ETH breaks down for finite $t'$ and $V'$, leading to a lack 
of thermalization~\cite{RigolPhysRevLett.103.100403}. An interesting question to ask is how this range of non-thermal
parameters varies with system size is increases. Does the range where the ETH holds increase to all
non-zero $t'$ and $V'$ as one approaches the thermodynamic limit? 
Initial numerical evidence on gapped systems supported that the ETH works
better for larger system sizes~\cite{RigolSantosPhysRevA.82.011604}. Based on analysis of scaling of the participation
ratio of eigenstates in an operator's eigenbasis, it was argued that indeed the ETH 
should be valid for any non-integrable parameters in the thermodynamic limit~\cite{NeuenhahnPhysRevE.85.060101}.

Another interesting question is whether or not numerical evidence shows that the ETH is valid 
in the {\em strong sense} as described in Sec. \ref{subsec:StatementOfETHI}. Numerical evidence
on lattice systems appears to support that it actually does~\cite{KimPhysRevE.90.052105}.
By explicitly searching for eigenstates where
expectation values are ``outliers": far away from their mean value, one can analyze how these scale
with system size. The behavior of the most extreme outliers as a function of system size, gives strong
numerical evidence that even these obey the ETH. 

Is the random matrix motivation for the ETH of \Sec{sec:RelRandMatrices} nothing more than a happy coincidence,
or is this the actual scenario for which it comes about? Aside from much earlier numerical work analyzing
the statistics of energy eigenstates and eigenvalues~\cite{FeingoldPhysRevA.34.591}, one can look at the ratio fluctuation sizes
of off diagonal matrix elements compared to diagonal elements. When compared with the results
of random matrix theory, the results agree very well for large enough systems~\cite{MondainiPhysRevE.96.012157}.

There is further evidence~\cite{geraedts2016many} that the connection of interacting systems to random matrix theory 
is the underlying reason for the validity of the ETH. This comes about by studying periodically driven {\em Floquet} 
systems~\cite{DAlessioAPolkovnikov2013,PhysRevE.90.012110,PonteChandran2015}. Ref. \cite{geraedts2016many} studied
systems that could either be in a many body localized, or in a thermal phase. They studied the {\em entanglement
spectrum} of such systems. In this case, they did so by tracing over half of their system and then considered the
logarithm of their resultant reduced density matrix. Regarding this as a kind of {\em entanglement Hamiltonian}, its
spectrum could be studied numerically. In the case of no driving, although they considered half of the system, 
they still found good agreement with the ETH in the thermal phase, supporting a strong version of it~\cite{garrison2015does}.
On the other hand, for periodic driving in the thermal phase, such systems heat up to infinite temperature~\cite{PonteChandran2015},
where ETH would then predict a trivial entanglement Hamiltonian equal to zero. Of course we expect corrections
to this, and these can be studied numerically. What is quite interesting about these corrections, is that they
also agree quite well with random matrix theory.

%% file: coldatoms.tex
\section{Relation to experiment}

\Sec{sec:routes} gave routes to understanding the origin of statistical
mechanics. It is not clear why any one but a dyed-in-the-wool theorist would care why it works,
as it seems so clear that it does. But as we have seen, isolated integrable systems are not expected
to produce systems that properly thermalize. Then again, quantum isolated integrable systems
would seem very hard to produce experimentally. 

The situation has changed dramatically over the last decade or so. There are now many experimental groups investigating
the properties of atoms trapped and isolated from the outside world at very low temperatures, down to picoKelvin.
These gases are typically quite dilute with a number of density of order $10^{14}/cm^3$. 
They are confined through magnetic or optical means
and the confinement can be of many forms, such as a harmonic potential,
or optical lattice~\cite{levin2012ultracold,langen2015ultracold}.
These isolated systems can have coherence times of seconds, while typical relaxation times are often in the millisecond
range~\cite{hofferberth2007non}, meaning that questions about isolated systems, that were purely theoretical twenty years ago, 
can now be tested experimentally. Furthermore control parameters,
such as magnetic field can be used to tune the two-body interaction via the Feschbach resonance~\cite{ChinRevModPhys.82.1225}.

In particular, the difference between integrable and non-integrable systems is very clear when
starting the system out of equilibrium, similar to the scenario considered in the introduction. A number of
one dimensional experimental systems are equivalent to integrable models~\cite{kinoshita2006quantum}, 
and very long lived momentum oscillations are observed.
In contrast to such integrable models, which have very long lived non-equilibrium behavior, higher dimensional situations
thermalize very rapidly~\cite{kinoshita2006quantum}. 
But although many aspects of these experiments are well described by theory, it is
difficult to produce energy eigenstates in order to directly test the ETH. These experiments do show that
isolated non-integrable quantum systems placed out of equilibrium,
do approach the results predicted from statistical mechanics, on a time scale far smaller
than their decoherence time. 

A general lesson that has been learnt from these kinds of experiments is that many of
the traditional approaches to understanding quantum statistical mechanics not only fail the theoretical
litmus test described in \Sec{subsec:LitmusTest}, 
but also fail to predict real experiments: There really is different behavior seen based on integrability,
and not all isolated systems do thermalize over experimentally important timescales.

In fact not only is thermalization not seen experimentally in integrable models, but also in some random disordered systems
that are believed to exhibit ``Many Body Localization"~\cite{choi2016exploring}, mentioned in \Sec{sec:ETH}. 
Not only can observables such as density and momentum distribution be measured, but quantities related
to entropy. In an optical lattice experiment~\cite{kaufman2016quantum}, with 6 rubidium atoms,
the second order R\'enyi entropy was determined by an ingenious method of interfering two copies
of the system. The growth of entanglement could be assessed throught this R\'enyi entropy, and
this entropy grew and saturated as predicted from ensemble statistical mechanical calculations.
Thus despite the small system size, thermalization was evident.

These experiments have also allowed the probing of systems that are much larger than those that
can be analyzed numerically, yet they still are quite precise, are highly tunable, and allow the measurement of many
local observables. In this way, they act as an intermediary between solid state experiments that
lack this precision, and numerics, which cannot attain such system sizes.

%% file: conclusions.tex
\section{Conclusions}
\label{sec:conclusions}

The Eigenstate Thermalization Hypothesis is considered by most researchers now to be the major conceptual tool in 
understanding how quantum mechanics leads to thermalization. Among other things, it allows is to understand 
how the time averages of measurements give rise to
the laws of statistical mechanics. In order to make headway, it makes sense to look at an isolated system
where outside sources cannot influence the dynamics. The most basic statistical mechanical concept in 
that case, is the microcanonical ensemble. 
The microcanonical ensemble considers energy states withing a shell of width $\Delta E$, of which there are
normally very many in the thermodynamic limit, even with very small $\Delta E$. 
The assumption of statistical mechanics is that time averages of
observables are given by expectation values averaged over all states within $\Delta E$. 
The ETH says that $\Delta E$ can be taken to be so narrow as to include {\em just a single energy}. 

An intuitive way of understanding the ETH, is to think of dividing the system into a smaller and larger region.
Even for an energy eigenstate, you can then think of the larger region as acting as a 
thermal ``bath" for the smaller region. 
The ETH says that for the system in any energy eigenstate, measurements
on only the smaller the region are equivalent to a system at the appropriate equivalent temperature. 
As discussed in this review, this will not work for an integrable system, for reasons that are
similar to the explanation in classical mechanics. In this sense, the ETH serves a similar purpose
as ergodicity, in connecting the microscopic dynamics to statistical mechanics.

The underlying explanation for the ETH appears to have to do with the relationship between non-integrable
quantum systems and random matrix theory. At a detailed microscopic level, at unimaginably small energy
scales, the energy level spacing statistics fit very well with random matrix models. And even the energy
eigenstates show a strong connection with random matrices. The ETH is a consequence of
this random matrix description, a statement that is
supported by detailed numerics for large enough systems~\cite{MondainiPhysRevE.96.012157}.

Aside for the current focus of the ETH in condensed matter and cold
atom systems, the ETH is now been used increasingly in the study of quantum
gravity, wormholes and firewalls. For example the ``ER=EPR" conjecture
of Maldacena and Susskind~\cite{maldacena2013cool} considers how
two black holes connected by a wormhole are related to the entanglement
between them. This has been analyzed using the ETH~\cite{marolf2013gauge}.
In another work, it has been proposed that the ETH can be applied to
the metric in quantum gravity~\cite{khlebnikov2014locality}.
As it turns out, understanding the microscopic structure of eigenstates
is important in a lot of applications.

It might seem odd that a statement about energy eigenstates for macroscopic systems would be
so helpful conceptually in understanding thermalization. The time it would take to prepare
a system in such an eigenstate is proportional to $\exp(S(E))$, which makes it completely inaccessible
in most experimental circumstances. But to check for ergodicity in a classical system also requires unattainable measurement
times, yet it is still crucial in understanding thermalization in that case. In both
the quantum and classical cases, these sorts of concepts allow one to distinguish behavior that
can be measured experimentally, such as the ability to thermalize. And in the quantum case
the structure is perhaps even richer, allowing for the existence of exotic systems that 
exhibit Many Body Localization~\cite{nandkishore2015many}, 
as briefly mentioned earlier. These
will likely continue to yield interesting new phenomena. And there are likely to be other
still undiscovered kinds of thermalization behavior that exist. For example, the idea of ``Quantum Disentangled Liquids"
has been proposed, where a system consisting of two different kinds of particles is 
only partially thermalized~\cite{grover2014quantum}. There are many reasons to believe that the 
recent focus in this area will continue to lead to many new surprises.

%% file: acknowledgments.tex
\section{Acknowledgements}

The author wishes to thank Richard Montgomery and Sergei Tabachnikov for an illuminating exchange 
on rigorous results for classical Sinai Billiards.
This research was supported by the Foundational Questions Institute \href{https://fqxi.org}{fqxi.org}.